\documentclass[preprint,12pt,authoryear]{elsarticle}

\usepackage{amssymb}
\usepackage{amsmath}
\usepackage{subcaption}
\DeclareMathOperator{\tr}{tr}
\journal{Journal of the Physics and Mechanics of Solids}

\begin{document}

\begin{frontmatter}

%% Title, authors and addresses

%% use the tnoteref command within \title for footnotes;
%% use the tnotetext command for theassociated footnote;
%% use the fnref command within \author or \address for footnotes;
%% use the fntext command for theassociated footnote;
%% use the corref command within \author for corresponding author footnotes;
%% use the cortext command for theassociated footnote;
%% use the ead command for the email address,
%% and the form \ead[url] for the home page:
%% \title{Title\tnoteref{label1}}
%% \tnotetext[label1]{}
%% \author{Name\corref{cor1}\fnref{label2}}
%% \ead{email address}
%% \ead[url]{home page}
%% \fntext[label2]{}
%% \cortext[cor1]{}
%% \address{Address\fnref{label3}}
%% \fntext[label3]{}

\title{Thermodynamic theory of crystal plasticity: formulation and application to polycrystal fcc copper}

%% use optional labels to link authors explicitly to addresses:
%% \author[label1,label2]{}
%% \address[label1]{}
%% \address[label2]{}

\author[ad1]{Charles K. C. Lieou}
\author[ad1,ad2]{Curt A. Bronkhorst\corref{cor1}}
\cortext[cor1]{Corresponding author}
\ead{cbronkhorst@wisc.edu}
\address[ad1]{Theoretical Division, Los Alamos National Laboratory, Los Alamos, NM 87545, USA\fnref{laur}}
\address[ad2]{Department of Engineering Physics, University of Wisconsin-Madison, Madison, WI 53706, USA}
\fntext[laur]{LA-UR release number: LA-UR-19-31861}

\begin{abstract}
%% Text of abstract
We present a thermodynamic description of crystal plasticity. Our formulation is based on the Langer-Bouchbinder-Lookman thermodynamic dislocation theory (TDT), which asserts the fundamental importance of an effective temperature that describes the state of configurational disorder and therefore the dislocation density of the crystalline material. We extend the TDT description from isotropic plasticity to crystal plasticity with many slip systems. Finite-element simulations show favourable comparison with experiments on polycrystal fcc copper under uniaxial compression, tension, and simple shear. The thermodynamic theory of crystal plasticity thus provides a thermodynamically consistent and physically rigorous description of dislocation motion in crystals. We also discuss new insights about the interaction of dislocations belonging to different slip systems.

\end{abstract}

\begin{keyword}
%% keywords here, in the form: keyword \sep keyword
Constitutive behavior \sep Crystal plasticity \sep Copper \sep Finite-element simulation \sep Taylor-Quinney coefficient
%% PACS codes here, in the form: \PACS code \sep code

%% MSC codes here, in the form: \MSC code \sep code
%% or \MSC[2008] code \sep code (2000 is the default)

\end{keyword}

\end{frontmatter}

%% \linenumbers

%% main text
\section{Introduction}
\label{sec:1}

The topic of representation of the mechanics and especially thermodynamics of dislocation motion and interaction remain areas of strong scientific and technological interest \citep[e.g.,][]{GurtinBook_2010,KubinBook_2013,BulatovBook_2007,ClaytonBook_2011}. In particular, significant questions and opportunities remain in understanding and quantifying material hardening, dislocation structure development, and deformation energy partitioning. Although challenges remain in experimental quantification of dislocation behavior, new lower length scale techniques offer important new opportunities to gather physical insight on dislocation behavior \citep[e.g.,][]{Pathak_2015,Thevamaran_2016,Pathak_2012,Xue_2017,Brown_2017,Upadhyay_2014,Gigax_2019}. Of equal importance are the many contributions made to define not only the kinematics of dislocation motion but also the energy states within thermodynamically consistent theories and computational frameworks \citep[e.g.,][]{Berdichevsky_2006,Berdichevsky_2017,Berdichevsky_2018,Berdichevsky_2018a,Berdichevsky_2019,Berdichevsky_2019a,anand_2015,Le_2018b,Le_2019,Hochrainer_2016,Levitas_2015b,Arora_2019,Po_2019,Chowdhury_2019,Jiang_2019,nietofuentes_2018,Jafari_2017,Shizawa_2001,RiveraDiazdelCastillo_2012,langer_2010,langer_2015,Roy_2005,Roy_2006,Acharya_2010}. Recent work \citep{Berdichevsky_2019} has pointed out statements of discouragement from the community in relationship to the difficulty of understanding and quantifying work hardening in metallic materials. Although difficult, the topic continues to be one of high economic and strategic value requiring models which are dedicated to specific materials \citep{Berdichevsky_2019}. Due to cumulative history effects and structural evolution, quantifying the details of the dislocation (and deformation twinning) mediated plasticity will be critical in predicting the ductile based damage of materials \citep{Boyce_2014,Boyce_2016}. Especially with regards to material damage, the ability to successfully predict when and where materials fail will require theories specified by material and properly account for energy and thermodynamic coupling between the different deformation mechanisms.

Early work in crystal plasticity theory had as a focus the structural representation of materials and the influence of the kinematics of discrete close-packed slip systems on crystallographic rotation and the development of crystallographic texture with deformation \citep[e.g.,][]{Hill_1972,Asaro_1977,Asaro_1983a, Asaro_1983b,Asaro_1985}. This large-deformation kinematical framework for models was an important element of enabling the prediction of crystallographic texture development \citep{Harren_1989a,Harren_1989b,Wu_1991,Bassani_1991,Bassani_1993,bronkhorst_1992,kalidindi_1992}. The crystallographic rotations had a close relationship with the imposed mode of deformation and dependence upon material type. The imprinting of crystallographic symmetries on macro-scale behaviors via granular rotations has significant implication for component manufacturing and material processing. This continues to be a very important area of application. The use of X-ray diffraction techniques supplementing mechanical testing enabled the study of preferred orientation of materials during large deformation and provided important experimental data to compare with simulations using these classical single crystal theories and homogenization techniques \citep{BungeBook_1982,Kocks_1989}. The years of work in developing homogenization strategies for engineering scale simulations of large deformation problems also inferred the importance of intergranular interactions \citep[e.g.,][]{VanHoutte_2005,VanHoutte_1982,VanHoutte_1988,Lebensohn_1993,bronkhorst_1992,kalidindi_1992,Lebensohn_2001}. Incorporating rate dependence in deformation also provided a natural selection process for dominant slip systems and remains an important element of most models \citep{Peirce_1982,Peirce_1983,Asaro_1985}. The models discussed here established the early kinematical and computational foundations but did not comprehensively account for energy considerations or thermodynamics of plastic processes.

With an eye towards greater abilities to predict the deformation response of metallic materials, there has been a systematic drive towards greater physical basis of continuum crystal models. Although dislocation motion is inherently discrete at the length scale of the Burgers vector, there have been many critical advances in continuum dislocation mechanics theories \citep{Arsenlis_1999,Arsenlis_2002,Gurtin_2000,Gurtin_2005,Acharya_2004,Anand_2005,Busso_1996,Busso_2000,Aifantis_1992,Fleck_1993,Fleck_1997,Zhu_1995,Gerken_2008,Mayeur_2011,Mayeur_2015} employing high level dislocation mechanics representations. More recently, theories for advanced representation of dislocation interaction and reactions have been formed and quantified using techniques of molecular dynamics and discrete dislocation dynamics \citep[e.g.,][]{KubinBook_2013,BulatovBook_2007,Bulatov_2017}. Given the relative simplicity of face-centered-cubic structures, advances in continuum representation of these materials have been more straighforward relative to BCC and HCP structures. A key advancement has been the move from a scalar representation of forest hardening to that of tensorial representation of specific slip system interactions and reactions \citep{Madec_2002,Madec_2003,Devincre_2006,Devincre_2008,Grilli_2018,dequiedt_2015,hansen_2013}. The work of \citet{hansen_2013} in particular includes non-mobile populations of dislocations which are sources of internal stored energy. In fact, some work suggests that dislocation sub-cell development is a result of lowering overall system energy \citep[e.g.,][]{Hansen_2010} and these continuum formulations have been used to study dislocation structural development during deformation at high resolutions \citep{dequiedt_2015,Grilli_2018,Arora_2019}.  

It has been known for some time that dislocation mediated plasticity is energy dissipative and a significant increase of temperature is possible \citep{farren_1925,taylor_1934}, especially when deformation is large and there is inadequate time for thermal transport to dissipate the energy. The proportion of plastic work transformed to thermal energy remains a very important area of research and is of technological interest to the processing of materials and material energy dissipation applications. For many types of loading conditions, the energy state of the material is an important state variable and the energy partitioned for dislocation state evolution must be considered \citep[e.g.,][]{Feng_2019,Wu_2003,mourad_2017,luscher_2018,Barton_2005,Feng_2019a,Vogler_2008,Levitas_2006,Levitas_2015b}. Despite more recent experimental work found in literature \citep{marchand_1988,Hartley_1987,duffy_1992,rittel_2012,rittel_2017} thermometry remains challenging to perform accurately and more experimental work is required. This body of work strongly suggests that the Taylor-Quinney coefficient \citep{taylor_1934} is not a constant value as is commonly assumed \citep[e.g.,][]{bronkhorst_2016} but can evolve significantly in magnitude with deformation. Recent work also suggests that the Taylor-Quinney coefficient differs in magnitude between slip and twinning processes \citep{Kingstedt_2019}. Together with the complexity of continuum descriptions of the mechanics of dislocation motion and interactions, theoretical development is also an active research area \citep{zehnder_1991,rosakis_2000,benzerga_2005,longere_2008a,longere_2008b,stainier_2010,zaera_2013,anand_2015,luscher_2018,nietofuentes_2018}. 

In the context of high deformation rate loading, we have proposed a novel thermodynamic description of structural evolution taking place during deformation \citep{lieou_2018,lieou_2019}. This approach is based upon recent work to partition energy and entropy elements to both atomic vibrational and atomic disorder states of matter \citep{langer_2010,langer_2015}. This general theory based upon atomic scale principles then provides a disciplined thermodynamic framework to describe structural features such as dislocations, dislocation subcells, grain boundaries, and so on. In our prior work \citep{lieou_2018,lieou_2019} and other recent applications of this theory \citep{langer_2016,langer_2017a,langer_2017b,le_2018a,Le_2018b,Le_2019} the atomic disorder component or dislocation, subcell, grain boundary representation was done only through isotropic and scalar representations of these states. There was but one dislocation density and grain boundary density measure used in the theory and subsequent simulations of experimental results. Here we advance upon the work in \citep{lieou_2018,lieou_2019} by applying this general thermodynamic approach to an advanced continuum theory of single crystal behavior for FCC systems. In this way, we significantly advance the description of the disorder (structure) partition component applied to specific crystallographic structure and begin to address the challenging questions raised above \citep{Berdichevsky_2006,Berdichevsky_2017,Berdichevsky_2018,Berdichevsky_2018a,Berdichevsky_2019,Berdichevsky_2019a,anand_2015,Le_2018b,Le_2019,Hochrainer_2016,Levitas_2015a,Arora_2019,Po_2019,Chowdhury_2019,Jiang_2019,nietofuentes_2018,Jafari_2017,Shizawa_2001,RiveraDiazdelCastillo_2012,langer_2010,langer_2015,Roy_2005,Roy_2006,Acharya_2010}. 

This paper is structured as follows. In Sec.~\ref{sec:2}, we present a crystal-plasticity formulation of the TDT, appropriately accounting for finite deformation and the resolution into different slip planes. We briefly discuss the computational method in Sec.~\ref{sec:3}. We validate our thermodynamic crystal-plasticity framework by direct comparison to a set of uniaxial compression experiments on copper in Sec.~\ref{sec:4}, and validate model results against simple shear and uniaxial tension in Sec.~\ref{sec:5}. We discuss some outstanding problems in Sec.~\ref{sec:6} and conclude the paper in Sec.~\ref{sec:7}. The nomenclature and quantities used within are given below in Table \ref{tab:symbols}.

\begin{table}
\scriptsize
\begin{center}
\caption{\label{tab:symbols}List of symbols}
\begin{tabular}{ll}
\hline
Symbol & Definition or meaning \\
\hline\hline
$\mathbf{F}$, $\mathbf{F}^e$, $\mathbf{F}^p$ & Total, elastic, and plastic deformation gradients \\
$\mathbf{L}$, $\mathbf{L}^e$, $\mathbf{L}^p$ & Total, elastic, and plastic velocity gradients \\
$\dot{\gamma}^{\alpha}$ & Resolved plastic strain rate on slip system $\alpha$ \\
$\mathbf{s}^{\alpha}$, $\mathbf{m}^{\alpha}$ & Unit slip direction vector and normal to slip system $\alpha$ \\
$\mathbf{\sigma}$ & Cauchy stress tensor \\
$\mathbf{S}$ & Second Piola-Kirchhoff stress tensor \\
$\mathbf{C}$ & Anisotropic fourth-order tensor of elastic constants \\
$C_{11}$, $C_{12}$, $C_{44}$ & Independent crystallographic moduli for fcc lattice \\
$\mu$ & Shear modulus \\
${\cal W}$ & Jacobian matrix of stress versus strain \\
$U_{\text{tot}}$, $S_{\text{tot}}$ & Total energy and entropy density \\
$U_K$, $S_K$ & Kinetic-vibrational (thermal) energy and entropy density \\
$U_C$, $S_C$ & Configurational energy and entropy density \\
$U_D$, $S_D$ & Dislocation energy and entropy \\
$U_1$, $S_1$ & Residual configurational energy and entropy density \\
$F_C$ & Configurational free energy density \\
$e_D$ & Dislocation line energy \\
$\chi$ & Effective temperature \\
$\theta$, $T$ & Thermal temperature (in units of energy and Kelvin) \\
$\rho^{\alpha}_{\text{ss}}$ & Steady-state dislocation density on slip system $\alpha$ \\
$\chi_{\text{ss}}$ & Steady-state effective temperature (in units of $e_D$) \\
$c_p$ & Specific heat capacity \\
$\kappa_{\rho}^{\alpha}$ & Dislocation storage rate \\
$\kappa_{\chi}$ & Effective temperature increase rate \\
$a$ & Minimum separation between dislocations \\
$b$ & Burgers vector \\
$t_0$ & Atomic time scale \\
$\alpha_T$ & Stress scale parameter \\
$\tau^{\alpha}$ & Resolved shear stress on slip system $\alpha$ \\
$s^{\alpha}_T$ & Slip resistance on slip system $\alpha$ \\
$\rho^{\alpha}$ & Dislocation density corresponding to slip system $\alpha$ \\
$l^{\alpha}$ & Dislocation mean free path on slip system $\alpha$ \\
$t^{\alpha}$ & Dislocation depinning time on slip system $\alpha$ \\
$T_P$ & Dislocation depinning barrier (in units of Kelvin) \\
$a^{\alpha \beta}$, $f^{\alpha \beta}$ & Dislocation interaction tensors \\
$d^{\alpha \beta}$ & Slip interaction tensor \\
$e_N$ & Interaction energy scale between dislocations on different slip systems \\
$k_c$, $k_{nc}$ & Mean free path parameters \\
\hline
\end{tabular}
\end{center}
\end{table}

%%%%% Section 2 %%%%%

\section{Thermodynamic theory of dislocation plasticity: a crystal-plasticity formulation}
\label{sec:2}

\subsection{Crystal plasticity and deformation kinematics}
\label{sec:2_1}

As in conventional crystal plasticity formulations, the deformation gradient $\mathbf{F}$ can be decomposed \citep{Kroner_1960} into elastic and plastic contributions, $\mathbf{F}^e$ and $\mathbf{F}^p$:
\begin{equation}
 \mathbf{F} = \mathbf{F}^e \mathbf{F}^p .
\end{equation}
Plastic velocity gradient is derived from the plastic shear strain rate on close-packed dislocation slip systems indexed by $\alpha$:
\begin{equation}\label{eq:L_p}
 \mathbf{L}^p \equiv \dot{\mathbf{F}}^p ( \mathbf{F}^p)^{-1} = \sum_{\alpha} \dot{\gamma}^{\alpha} \mathbf{s}^{\alpha} \otimes \mathbf{m}^{\alpha} ,
\end{equation}
where $\mathbf{s}^{\alpha}$ is the unit slip direction vector, $\mathbf{m}^{\alpha}$ is the unit normal to the slip plane vector $\alpha$, and $\dot{\gamma}^{\alpha}$ is the plastic strain rate on slip system ${\alpha}$, all defined in the reference configuration. If $\mathbf{S}$ is a stress tensor of the second Piola-Kirchhoff type, defined by
\begin{equation}\label{eq:PK2}
 \mathbf{S} \equiv J \, (\mathbf{F}^e)^{-1} \boldsymbol{\sigma} (\mathbf{F}^e)^{-T} ,
\end{equation}
where $\boldsymbol{\sigma}$ is the Cauchy stress tensor, and $J = \det \mathbf{F} = \det \mathbf{F}^e$ is the Jacobian of the transformation between the reference and deformed configurations (the second inequality holds because we assume plastic incompressibility, $\det \mathbf{F}^p = 1$), then the resolved shear stress $\tau^{\alpha}$ on slip system $\alpha$ is given by
\begin{equation}
 \tau^{\alpha} = (\mathbf{F}^e)^T \mathbf{F}^e \mathbf{S} \colon ( \mathbf{s}^{\alpha} \otimes \mathbf{m}^{\alpha}) \approx \mathbf{S} \colon ( \mathbf{s}^{\alpha} \otimes \mathbf{m}^{\alpha}) ,
\end{equation}
where the approximation follows because elastic strains for conditions of loading considered in this work are small. The resolved shear stress thus defined provides a simple expression for the total plastic work rate:
\begin{equation}\label{eq:W_p}
 \boldsymbol{\sigma} \colon \mathbf{F}^e \mathbf{L}^p (\mathbf{F}^{e})^{-1} = \sum_{\alpha} \tau^{\alpha} \dot{\gamma}^{\alpha} ,
\end{equation}
the right-hand side being the sum of the plastic work rates on all slip systems.

\subsection{Stress response}
\label{sec:2_2}

The second Piola-Kirchhoff stress tensor $\mathbf{S}$ is related to the elastic part of the deformation gradient $\mathbf{F}^e$ by
\begin{equation}
 \mathbf{S} = \dfrac{1}{2} \mathbf{C} \left( (\mathbf{F}^e)^T \mathbf{F}^e - \mathbf{I} \right) ,
\end{equation}
where $\mathbf{C}$ is the anisotropic fourth-order tensor of elastic constants. For cubic crystallographic systems, this tensor is composed of three independent moduli in $\mathit{C}_{\text{11}}$, $\mathit{C}_\text{{12}}$, and $\mathit{C}_\text{{44}}$.

\subsection{Thermodynamics and state variable evolution}
\label{sec:2_3}

As in many conventional theories of dislocation plasticity \citep[e.g.,][]{cheong_2004,hansen_2013,anand_2015,dequiedt_2015,zecevic_2015,knezevic_2018,bronkhorst_2019}, the relevant internal state variable is the dislocation density $\rho^{\alpha}$ on the different slip systems $\alpha$. What distinguishes the present approach from conventional theories is that the evolution of the dislocation density are derived from energetic and entropic considerations alone, subject to the constraints of the first and second laws of thermodynamics. Our derivation here is largely parallel to that described in simpler and isotropic set-ups \citep[e.g.,][]{langer_2010,langer_2015,lieou_2018,lieou_2019}, but is now adapted to incorporate finite deformation and different slip systems with some initial accounting for dislocation interactions.

Let $U_{\text{tot}}$ and $S_{\text{tot}}$ denote the total energy and entropy density per unit volume of the material, each being a sum of kinetic-vibrational (i.e., thermal motion of atoms) and configurational (i.e., positions of atoms, defects, etc.) contributions:
\begin{equation}
 U_{\text{tot}} = U_K + U_C; \quad S_{\text{tot}} = S_K + S_C .
\end{equation}
$U_C$ and $S_C$ are clearly functions of relevant internal state variables and, in particular, of the dislocation densities $\rho^{\alpha}$, so that
\begin{eqnarray}
 U_C (S_C, \rho^{\alpha} ) &=& U_D (\rho^{\alpha}) + U_1 (S_1) ; \\
 S_C (U_C, \rho^{\alpha} ) &=& S_D (\rho^{\alpha}) + S_1 (U_1) .
\end{eqnarray}
where $U_D$ and $S_D$ denote contributions from dislocations, and $U_1$ and $S_1$ are the residual configurational energy and entropy densities for all other configurational degrees of freedom (e.g., impurities, point defects). An important quantity to our theoretical development is the effective temperature defined as
\begin{equation}
 \chi \equiv \dfrac{\partial U_C}{\partial S_C} ,
\end{equation}
which quantifies the atomic disorder in the material relative to the perfect crystal.

Before specifying the dependence of $U_C$ and $S_C$ on the dislocation densities $\rho^{\alpha}$, let us first examine the consequences of the laws of thermodynamics. The first law of thermodynamics says that
\begin{eqnarray}\label{eq:first_law}
 \nonumber \dot{U}_{\text{tot}} &=& \boldsymbol{\sigma} \colon \mathbf{L} = \dot{U}_C + \dot{U}_K \\ &=& \chi \dot{S}_C + \left( \dfrac{\partial U_C}{\partial t} \right)_{S_C, \rho^{\alpha}} + \sum_{\alpha} \left( \dfrac{\partial U_C}{\partial \rho^{\alpha}} \right)_{S_C} \dot{\rho}^{\alpha} + \theta \dot{S}_K ,
\end{eqnarray}
where $\theta = k_B T$ is the ordinary, thermal temperature in energy units ($k_B$ being the Boltzmann constant), and $\mathbf{L} = \dot{\mathbf{F}} \mathbf{F}^{-1}$ is the velocity gradient, which can be separated into elastic and plastic contributions:
\begin{equation}
 \mathbf{L} = \mathbf{L}^e + \mathbf{F}^e \mathbf{L}^p ( \mathbf{F}^e )^{-1} ,
\end{equation}
with $\mathbf{L}^e = \dot{\mathbf{F}}^e (\mathbf{F}^e )^{-1}$, and $\mathbf{L}^p$ given above in Eq.~\eqref{eq:L_p}. Observe now that
\begin{equation}
 \left( \dfrac{\partial U_C}{\partial t} \right)_{S_C, \rho^{\alpha}} = \boldsymbol{\sigma} \colon \mathbf{L}^e
\end{equation}
because deformation at fixed configurational state is elastic by definition. Thus
\begin{equation}\label{eq:first_law2}
 \boldsymbol{\sigma} \colon \mathbf{F}^e \mathbf{L}^p ( \mathbf{F}^e )^{-1} = \chi \dot{S}_C + \sum_{\alpha} \left( \dfrac{\partial U_C}{\partial \rho^{\alpha}} \right)_{S_C} \dot{\rho}^{\alpha} + \theta \dot{S}_K ,
\end{equation}
Turn now to the second law of thermodynamics, which states that
\begin{equation}\label{eq:second_law}
 \dot{S}_{\text{tot}} = \dot{S}_C + \dot{S}_K \geq 0 .
\end{equation}
Multiplying this by $\chi$, and eliminating $\dot{S}_C$ using Eq.~\eqref{eq:first_law2} above, we get
\begin{equation}
 \boldsymbol{\sigma} \colon \mathbf{F}^e \mathbf{L}^p ( \mathbf{F}^e )^{-1} - \sum_{\alpha} \left( \dfrac{\partial U_C}{\partial \rho^{\alpha}} \right)_{S_C} \dot{\rho}^{\alpha} + (\chi - \theta ) \dot{S}_K \geq 0.
\end{equation}
According to the Coleman-Noll procedure \citep{coleman_1963}, each independently variable term must be non-negative in order for the overall inequality to hold. This follows automatically for the plastic work term (see Eq.~\eqref{eq:W_p} above). Thus we arrive at the constraints
\begin{equation}\label{eq:secondlaw_ineq}
 - \left( \dfrac{\partial U_C}{\partial \rho^{\alpha}} \right)_{S_C} \dot{\rho}^{\alpha} \geq 0; \quad (\chi - \theta ) \dot{S}_K \geq 0 .
\end{equation}
The second of these inequalities constrains the Taylor-Quinney factor which determines the amount of plastic power that goes into heating up the material, as has been discussed in \citet{lieou_2019}. The implications of the first inequality can be seen by writing $U_C (S_C, \rho^{\alpha}) = U_D (\rho^{\alpha}) + U_1 (S_1) = U_D (\rho^{\alpha}) + U_1 (S_C - S_D (\rho^{\alpha}))$, from which it follows that
\begin{equation}\label{eq:F_C_partial}
 \left( \dfrac{\partial U_C}{\partial \rho^{\alpha}} \right)_{S_C} = \dfrac{\partial U_D}{\partial \rho^{\alpha}} - \chi \dfrac{\partial S_D}{\partial \rho^{\alpha}} \equiv \dfrac{\partial F_C}{\partial \rho^{\alpha}} ,
\end{equation}
where
\begin{equation}\label{eq:F_C}
 F_C (\rho^{\alpha}) = U_D (\rho^{\alpha}) - \chi S_D (\rho^{\alpha})
\end{equation}
is the configurational free energy density. (The second term in the first equality in Eq.~\eqref{eq:F_C_partial} follows from a direct application of the chain rule of differential calculus to $U_1$ and the definition of $\chi$). Thus the first of Eq.~\eqref{eq:secondlaw_ineq} implies that deformation causes the material to dynamically minimize its configurational free energy. In addition, the material reaches its non-equilibrium steady state, i.e., $\dot{\rho}^{\alpha} = 0$, when $\partial F_C / \partial \rho^{\alpha} = 0$. This non-equilibrium steady state is possible when the rate of dislocation generation due to multiplication processes balance the rate of dislocation annihilation due to interaction processes. This means that the steady state dislocation densities $\rho^{\alpha}_{\text{ss}}$ can be obtained by locating configurational free energy minima which are generally a function of material state and loading conditions.

The entropy density $S_D$ can be computed by directly counting the microstates across slip systems (e.g., \citet{lieou_2018}); the result is
\begin{equation}\label{eq:S_D}
 S_D = \dfrac{1}{a} \sum_{\alpha} \left[ - \rho^{\alpha} \ln (a^2 \rho^{\alpha} ) + \rho^{\alpha} \right] .
\end{equation}
where the quantity $a$ is an atomic length scale, not necessarily the Burgers vector magnitude $b$. We shall specify the corresponding dislocation energy density $U_D$ shortly afterwards when we specify the interaction between different slip systems in Sec.~\ref{sec:2_4}.

Based on energetic considerations alone, the evolution of the dislocation density must be proportional to the rate at which input work is stored in newly formed dislocations, which in turn is proportional to the plastic work rate. This contrasts with conventional crystal plasticity theories, according to which the dislocation density evolves at a rate proportional to the absolute value of the slip rate of the corresponding slip system. Thus our proposed evolution equation for $\rho^{\alpha}$ is of the form
\begin{equation}\label{eq:rhodot}
 \dot{\rho}^{\alpha} = \dfrac{\kappa_{\rho}^{\alpha}}{a^2} \dfrac{ \tau^{\alpha} \dot{\gamma}^{\alpha}}{\mu} \left( 1 - \dfrac{\rho^{\alpha}}{\rho^{\alpha}_{\text{ss}}} \right) .
\end{equation}
Equation \eqref{eq:rhodot}, as written, explicitly assumes that dislocation creation on a given slip system is driven only by plastic slip on that system. This is the case for conventional dislocation theories as well \citep[e.g.,][]{cheong_2004,hansen_2013,anand_2015,dequiedt_2015,zecevic_2015,knezevic_2018,bronkhorst_2019}. Here, $\kappa_{\rho}^{\alpha}$ is a dynamic parameter that controls the hardening rate; $\mu$, the shear modulus and only relevant stress scale for deviatoric deformation, is inserted in the denominator for dimensional consistency.

The effective temperature $\chi$ follows an evolution equation of the form
\begin{equation}\label{eq:chidot}
 \dfrac{\dot{\chi}}{e_D} = \kappa_{\chi} \dfrac{\sum_{\beta} \tau^{\beta} \dot{\gamma}^{\beta}}{ \mu} \left(1 - \dfrac{\chi}{\chi_{\text{ss}}} \right) ,
\end{equation}
where $\kappa_{\chi}$ is a dimensionless parameter, and $\chi_{\text{ss}}$ is the steady-state effective temperature, which may be strain-rate dependent. Finally, if there is no heat exchange with the surroundings and conditions can be assumed to be adiabatic, the thermal temperature evolves according to
\begin{equation}\label{eq:Tdot}
 \dot{T} = \dfrac{\beta_T}{\bar{\rho}_M c_p} \sum_{\beta} \tau^{\beta} \dot{\gamma}^{\beta} ,
\end{equation}
where $\beta_T \approx \chi / \chi_{\text{ss}}$, the Taylor-Quinney coefficient, quantifies the fraction of plastic power converted into heat. $\bar{\rho}_M$ is the mass density of the material, and $c_p$ is the specific heat capacity of the material per unit mass. At fast loading rates, one can assume adiabatic deformation, and neglect the exchange of heat with the surroundings and the flow of heat across the material. At quasistatic loading rates, one can assume isothermal conditions and neglect the temporal evolution of $T$.

\subsection{Kinematics of dislocations and interaction between different slip systems}
\label{sec:2_4}

To complete our theoretical description we need an expression for the plastic strain rate $\dot{\gamma}^{\alpha}$ on slip system $\alpha$. This is given by the Orowan relation:
\begin{equation}\label{eq:orowan}
 \dot{\gamma}^{\alpha} = \rho^{\alpha} b v^{\alpha},
\end{equation}
where $v^{\alpha}$ is the average dislocation speed on slip system $\alpha$.

We regard dislocation motion as a stress-driven, thermally-activated depinning process that takes place on a time scale $t^{\alpha}$ given by
\begin{equation}
 \dfrac{1}{t^{\alpha}} = \dfrac{1}{t_0} \exp \left( - \dfrac{T_P}{T} e^{- \tau^{\alpha} / s_T^{\alpha} }\right) ,
\end{equation}
where $T$ is the temperature in units of Kelvin, and $T_P$ characterizes the depinning energy barrier through the product $k_B T_P$ with the Boltzmann constant $k_B$, and $s_T^{\alpha}$ the slip resistance on slip system $\alpha$. As is commonly done, if $l^{\alpha}$ denotes the mean-free path of a single dislocation in slip system $\alpha$ in the dislocation forest, one would expect that $v^{\alpha} = l^{\alpha} / t^{\alpha}$, so that
\begin{equation}
 \dot{\gamma^{\alpha}} = \dfrac{\rho^{\alpha} l^{\alpha} b}{t_0} \exp \left( - \dfrac{T_P}{T} e^{- \tau^{\alpha} / s_T^{\alpha} }\right) .
\end{equation}

In the scalar, isotropic version of the TDT, the expressions for the mean-free path and the slip resistance are relatively simple: $l = 1 / \sqrt{\rho}$ and $s_T \propto \sqrt{\rho}$, where we suppress the slip system index $\alpha$ because of its irrelevance. Also, the dislocation energy density $U_D$ discussed above in Sec.~\ref{sec:2_3} is linear in the dislocation density in the non-interacting approximation: $U_D = \dfrac{e_D \rho}{a}$. All of these, however, are places where interaction between slip systems could come into play in the present crystal-plasticity formulation. Specifically,

\begin{enumerate}

\item The slip resistance, or the Taylor stress, could couple the dislocation densities $\rho^{\alpha}$ corresponding to the different slip systems, for dislocation motion is impeded by the dislocation forest itself:
\begin{equation}\label{eq:taylorstress}
 s^{\alpha}_T = \alpha_T \mu \, b \, \sqrt{\sum_{\beta} a^{\alpha \beta} \rho^{\beta}},
\end{equation}
where $a^{\alpha \beta}$ is a dislocation interaction tensor. The scaling factor $\alpha_T$ is inserted in light of the uncertainty between the stress barrier for traditional theories documented in the literature \citep[e.g.,][]{hansen_2013}, and the stress barrier in the present theory.

\item The mean-free path of a dislocation in slip system $\alpha$ could depend on the dislocation densities in other slip systems of the dislocation forest:
\begin{equation}
 l^{\alpha} = \dfrac{1}{\sqrt{\sum_{\beta} d^{\alpha \beta} \rho^{\beta}}},
\end{equation}
where $d^{\alpha \beta}$ is the slip interaction tensor.

\item The dislocation formation energy $U_D$ could contain an interaction term that couples the different systems:
\begin{equation}\label{eq:U_D}
 U_D (\rho^{\alpha}) = \dfrac{1}{a} \left( \sum_{\alpha} e_D \rho^{\alpha} + \dfrac{b^2 e_N}{2} \sum_{\alpha} \sum_{\beta} f^{\alpha \beta} \rho^{\alpha} \rho^{\beta} \right) ,
\end{equation}
where $f^{\alpha \beta}$ is yet another dislocation interaction tensor, and $e_N$ is an interaction energy scale. With the entropy $S_D$ given in Eq.~\eqref{eq:S_D}, the steady-state dislocation energy density is given by
\begin{equation}\label{eq:rho_ss0}
 \rho_{\text{ss}}^{\alpha} = \dfrac{1}{a^2} \exp \left( - \dfrac{e_D + b^2 e_N \sum_{\beta} f^{\alpha \beta} \rho^{\beta}}{\chi} \right) .
\end{equation}
This expression implies the interesting possibility that an accumulation of dislocations in some slip planes could cause a reduction of the dislocation density in other slip planes.
\end{enumerate}
Possibilities 1 and 2 have direct counterparts in conventional dislocation theories such as theories of the Kocks-Mecking type \citep[e.g.,][]{mecking_1981}, and appear to give us better fits to the copper stress-strain data that the present manuscript is concerned with. Setting the dislocation interaction energy equal to zero gives a simple expression for the steady-state dislocation density:
\begin{equation}\label{eq:rho_ss}
 \rho_{\text{ss}}^{\alpha} = \dfrac{1}{a^2} e^{- e_D / \chi} .
\end{equation}
However, possibility 3 -- a non-trivial interaction energy between different slip systems -- remains an interesting prospect, which we intend to address in future work. From here onwards, we adopt units in which $e_D = 1$, and render $\chi$ dimensionless.

With these ingredients, the slip rate is given by
\begin{equation}\label{eq:gammadot_1}
 \dot{\gamma}^{\alpha}_{|\tau^{\alpha}| > 0} = \dfrac{\bar{\rho}^{\alpha}}{t_0 \sqrt{\sum_{\beta} d^{\alpha \beta} \bar{\rho}^{\beta}}} \exp \left[ - \dfrac{T_P}{T} e^{- \tau^{\alpha} / (\mu \sqrt{\sum_{\beta} a^{\alpha \beta} \bar{\rho}^{\beta}}) }\right] \equiv \dfrac{f (\tau^{\alpha}, \bar{\rho}^{\alpha})}{t_0},
\end{equation}
where $\bar{\rho}^{\alpha} \equiv b^2 \rho^{\alpha}$ is the rescaled, dimensionless dislocation density.

We now impose the requirement that upon stress reversal, $\tau^{\alpha} \rightarrow - \tau^{\alpha}$, the strain rate is also reversed: $\dot{\gamma}^{\alpha} \rightarrow - \dot{\gamma}^{\alpha}$. (This requirement was not necessary in polycrystalline plasticity calculations where we used the von Mises stress $\bar{s}$, which is always non-negative, in place of the resolved shear stress $\tau^{\alpha}$; see for example \cite{lieou_2018}.) Thus we subtract from Eq. \eqref{eq:gammadot_1} the term $f( - \tau^{\alpha}, \bar{\rho}^{\alpha})$ to preserve symmetry:
\begin{equation}\label{eq:gammadot_2}
\dot{\gamma}^{\alpha} = \dfrac{1}{t_0} [f( \tau^{\alpha}, \bar{\rho}^{\alpha}) - f( - \tau^{\alpha}, \bar{\rho}^{\alpha})] \equiv \dfrac{g (\tau^{\alpha}, \bar{\rho}^{\alpha})}{t_0} .
\end{equation}
This is necessary to ensure that dislocation slip is strictly dissipative. Since structural evolution is driven by plastic work rate as given in Eqs. (21) and (22), it remains physically correct with load reversal. Additionally, following \citet{langer_2010} and \citet{langer_2015}, Eq.~\eqref{eq:gammadot_1} for the slip rate allows us to compute the hardening parameter $\kappa_{\rho}^{\alpha}$, by considering the onset of strain hardening, when the resolved shear stress is roughly equal to the Taylor stress $s_T^{\alpha}$, but the dislocation density is still small. The conclusion is that $\kappa_{\rho}^{\alpha}$ can be written in the form
\begin{equation}
 \kappa_{\rho}^{\alpha} = \dfrac{\kappa_1^{\alpha}}{(\bar{\nu}^{\alpha})^2} ,
\end{equation}
where
\begin{equation}
 \bar{\nu}^{\alpha} \equiv \ln \left( \dfrac{T_P}{T} \right) - \ln \left[ \ln \left( \dfrac{\bar{\rho}^{\alpha}}{t_0 | \dot{\gamma}_0^{\alpha} | \sqrt{\sum_{\beta} d^{\alpha \beta} \bar{\rho}^{\beta} } } \right) \right] ,
\end{equation}
with
\begin{equation}
 \dot{\gamma}_0^{\alpha} \equiv (\mathbf{F}^e)^{-1} \mathbf{L} \mathbf{F}^e \colon ( \mathbf{s}^{\alpha} \otimes \mathbf{m}^{\alpha} )
\end{equation}
being the total shear rate resolved in slip plane $\alpha$.

%%%%% Section 3 %%%%%

\section{Computational method}
\label{sec:3}

The thermodynamic theory of crystal plasticity outlined above was implemented in the implicit branch of Abaqus \citep{abaqus_2014}, following the numerical integration scheme outlined in \citet{kalidindi_1992} and \citet{bronkhorst_1992}. We simulated a polycrystal fcc copper cube of lateral dimension 1 mm comprised of 1000 eight-node linear brick elements with full integration undergoing uniaxial compression, up to a true strain of unity. The simple computational model assumed each element was assigned an individual grain and periodic boundary conditions applied to the surfaces as in \citep{bronkhorst_2007}. The same polycrystal cube was also used for the simple shear and uniaxial tension simulations, to be described in Sec.~\ref{sec:5}.

The implicit finite-element method uses a Newton-type iteration scheme for computing estimates of the nodal displacement, which necessitates the computation of the Jacobian matrix
\begin{equation}\label{eq:jacobian_def}
 {\cal W} = \dfrac{\partial \vec{\boldsymbol{\sigma}}}{\partial \vec{\mathbf{e}}} ,
\end{equation}
where $\vec{\boldsymbol{\sigma}}$ and $\vec{\mathbf{e}}$ are vector representations of the Cauchy stress tensor $\boldsymbol{\sigma}$ and symmetric relative strain tensor $\mathbf{e}$, evaluated at time $t + \Delta t$ as follows:
\begin{equation}
 \vec{\boldsymbol{\sigma}} = \begin{bmatrix} \sigma_{11} (t + \Delta t) \\ \sigma_{22} (t + \Delta t) \\ \sigma_{33} (t + \Delta t) \\ \sigma_{12} (t + \Delta t) \\ \sigma_{13} (t + \Delta t) \\ \sigma_{23} (t + \Delta t) \end{bmatrix}; \quad \vec{\mathbf{e}} = \begin{bmatrix} e_{11} (t + \Delta t) \\ e_{22} (t + \Delta t) \\ e_{33} (t + \Delta t) \\ e_{12} (t + \Delta t) \\ e_{13} (t + \Delta t) \\ e_{23} (t + \Delta t) \end{bmatrix} .
\end{equation}
The relative strain tensor $\mathbf{e}$ is given by
\begin{equation}
 \mathbf{e} = \ln \mathbf{U}^t ,
\end{equation}
where $\mathbf{U}^t$ is the relative stretch obtained from the polar decomposition of the relative deformation gradient $\mathbf{F}^t$:
\begin{equation}\label{eq:Ft_def}
 \mathbf{F} (t + \Delta t) = \mathbf{F}^t \mathbf{F}(t) ; \quad \mathbf{F}^t = \mathbf{R}^t \mathbf{U}^t .
\end{equation}

To compute the Jacobian in Eq.~\eqref{eq:jacobian_def}, note first the relationship between the Cauchy stress tensor $\boldsymbol{\sigma} (t + \Delta t)$ and the second Piola-Kirchoff stress tensor $\mathbf{S} (t + \Delta t)$ at time $t + \Delta t$ are related by (c.f. Eq.~\eqref{eq:PK2} above)
\begin{equation}
 \boldsymbol{\sigma}(t + \Delta t) = \dfrac{1}{\det \mathbf{F}^e (t + \Delta t)}\mathbf{F}^e (t + \Delta t) \mathbf{S} (t + \Delta t) (\mathbf{F}^e)^T (t + \Delta t) ,
\end{equation}
so that, upon taking infinitesimal variation and suppressing the argument $t + \Delta t$,
\begin{equation}
 \delta \boldsymbol{\sigma} = \dfrac{1}{\det \mathbf{F}^e} \left[ \delta \mathbf{F}^e \mathbf{S} (\mathbf{F}^e)^T + \mathbf{F}^e \delta \mathbf{S} (\mathbf{F}^e)^T + \mathbf{F}^e \mathbf{S} \delta (\mathbf{F}^e)^T - \left( \mathbf{F}^e \mathbf{S} (\mathbf{F}^e)^T \right) \tr \left( \delta \mathbf{F}^e (\mathbf{F}^e)^{-1} \right) \right] . ~~~~~
\end{equation}
If we define the fourth-rank tensors ${\cal T}$ and ${\cal Q}$ as follows:
\begin{equation}
 {\cal T} \equiv \dfrac{\partial \mathbf{F}^e}{\partial \mathbf{e}}; \quad {\cal Q} \equiv \dfrac{\partial \mathbf{S}}{\partial \mathbf{e}},
\end{equation}
then the fourth-rank representation of the Jacobian defined in Eq.~\eqref{eq:jacobian_def} is given by
\begin{eqnarray}\label{eq:jacobian4}
 \delta \sigma_{ij} &\equiv& {\cal W}_{ijkl} \delta e_{kl} ; \\
 {\cal W}_{ijkl} &=& \dfrac{1}{\det \mathbf{F}^e} \left[ {\cal T}_{imkl} S_{mn} F^{e^T}_{nj} + F^e_{im} {\cal Q}_{mnkl} F^{e^T}_{nj} + F^e_{im} S_{mn} {\cal T}_{jnkl} - F^e_{im} S_{mn} F^{e^T}_{nj} ({\cal T}_{pqkl} F^{e^{-1}}_{qp} )\right] . ~~~~~
\end{eqnarray}
This can be converted to the second-rank form in Eq.~\eqref{eq:jacobian_def} by means of the transformation
\begin{equation}
 {\cal W} =
 \begin{bmatrix}
 {\cal W}_{1111} & {\cal W}_{1122} & {\cal W}_{1133} & ({\cal W}_{1112}+{\cal W}_{1121})/2 & ({\cal W}_{1113}+{\cal W}_{1131})/2 & ({\cal W}_{1123} + {\cal W}_{1132})/2 \\
 {\cal W}_{2211} & {\cal W}_{2222} & {\cal W}_{2233} & ({\cal W}_{2212}+{\cal W}_{2221})/2 & ({\cal W}_{2213}+{\cal W}_{2231})/2 & ({\cal W}_{2223} + {\cal W}_{2232})/2 \\
 {\cal W}_{3311} & {\cal W}_{3322} & {\cal W}_{3333} & ({\cal W}_{3312}+{\cal W}_{3321})/2 & ({\cal W}_{3313}+{\cal W}_{3331})/2 & ({\cal W}_{3323} + {\cal W}_{3332})/2 \\
 {\cal W}_{1211} & {\cal W}_{1222} & {\cal W}_{1233} & ({\cal W}_{1212}+{\cal W}_{1221})/2 & ({\cal W}_{1213}+{\cal W}_{1231})/2 & ({\cal W}_{1223} + {\cal W}_{1232})/2 \\
 {\cal W}_{1311} & {\cal W}_{1322} & {\cal W}_{1333} & ({\cal W}_{1312}+{\cal W}_{1321})/2 & ({\cal W}_{1313}+{\cal W}_{1331})/2 & ({\cal W}_{1323} + {\cal W}_{1332})/2 \\
 {\cal W}_{2311} & {\cal W}_{2322} & {\cal W}_{2333} & ({\cal W}_{2312}+{\cal W}_{2321})/2 & ({\cal W}_{2313}+{\cal W}_{2331})/2 & ({\cal W}_{2323} + {\cal W}_{2332})/2
 \end{bmatrix} .
\end{equation}
For small incremental stretch,
\begin{equation}
 \mathbf{e} = \ln \mathbf{U}^t \approx \mathbf{U}^t - 1 ,
\end{equation}
so that $\delta \mathbf{e} = \delta \mathbf{U}^t$, and hence
\begin{equation}\label{eq:TQmat_def}
 {\cal T} \approx \dfrac{\partial \mathbf{F}^e}{\partial \mathbf{U}^t}; \quad {\cal Q} \approx \dfrac{\partial \mathbf{S}}{\partial \mathbf{U}^t} .
\end{equation}
It suffices to compute these two quantities in order to evaluate the Jacobian ${\cal W}$. We list the steps below without proof; readers interested in the derivation may consult, for example, \citet{BalasubramanianThesis_1998}.

\begin{enumerate}

\item Compute the fourth-rank matrix
\begin{equation*}
 {\cal L}_{ijkl} = F^{e^T}_{ik} (t) U^t_{lm} (t + \Delta t) F^e_{mj} (t) + F^{e^T}_{im} (t) U^t_{mk} (t + \Delta t) F^e_{ij} (t) .
\end{equation*}

\item Let ${\cal C}^c_{ijkl}$ denote the anisotropic elastic modulus tensor in the crystal basis, and let $\mathbf{Q}$ denote the orthogonal rotation from the crystal basis to the global coordinates. Compute the anisotropic elastic modulus tensor in the global basis, ${\cal C}_{ijkl}$, as follows:
\begin{equation*}
 {\cal C}_{ijkl} = Q_{im} Q_{jn} Q_{kp} Q_{lq} {\cal C}^c_{mnpq} .
\end{equation*}
Use this result to calculate
\begin{equation*}
 {\cal D}_{ijkl} = \dfrac{1}{2} {\cal C}_{ijmn} {\cal L}_{mnkl} .
\end{equation*}

\item Let $\mathbf{N}^{\alpha} \equiv \mathbf{s}^{\alpha} \otimes \mathbf{m}^{\alpha}$ be the Schmid tensor for slip system $\alpha$, in the reference configuration. For each $\alpha$, compute
\begin{eqnarray}
 \nonumber {\cal G}^{\alpha}_{mnkl} &=& {\cal L}_{mpkl} N^{\alpha}_{pn} + N^{\alpha^T}_{mp} {\cal L}_{pnkl} ; \\
 \nonumber {\cal J}^{\alpha}_{ijkl} &=& \dfrac{1}{2} {\cal C}_{ijmn} {\cal G}^{\alpha}_{mnkl} .
\end{eqnarray}

\item Compute the matrix
\begin{equation*}
 B^{\alpha}_{ij} = \dfrac{\Delta t}{2} \dfrac{\partial \dot{\gamma}^{\alpha}(t)}{\partial \tau^{\alpha}} \left( N^{\alpha}_{ij} + N^{\alpha}_{ji} \right) .
\end{equation*}

\item Compute the fourth-rank matrices
\begin{eqnarray}
 \nonumber {\cal K}_{ijkl} &=& {\cal I}_{ijkl} + \sum_{\alpha} C^{\alpha}_{ij} B^{\alpha}_{kl} ; \\
 \nonumber {\cal Q}_{ijkl} &=& {\cal K}^{-1}_{ijmn} \left( {\cal D}_{mnkl} - \sum_{\alpha} \dot{\gamma}^{\alpha} \Delta t {\cal J}^{\alpha}_{mnkl} \right) .
\end{eqnarray}
In the first of these equations, ${\cal I}_{ijkl}$ is the fourth-rank identity matrix.

\item Compute the quantities
\begin{eqnarray}
 \nonumber R^{\alpha}_{ij} &=& B^{\alpha}_{kl} {\cal Q}_{klij} ; \\
 \nonumber {\cal T}_{ijkl} &=& R^t_{ik} \left( F^e_{lj} (t) - F^e_{lp} (t) \sum_{\alpha} \dot{\gamma}^{\alpha} \Delta t N^{\alpha}_{pj} \right) - R^t_{im} U^t_{mn} F^e_{np} (t) \sum_{\alpha} R^{\alpha}_{kl} N^{\alpha}_{pj} .
\end{eqnarray}

\item Finally, compute the Jacobian
\begin{equation*}
 {\cal W}_{ijkl} = \dfrac{1}{\det \mathbf{F}^e} \left[ {\cal T}_{imkl} S_{mn} F^{e^T}_{nj} + F^e_{im} {\cal Q}_{mnkl} F^{e^T}_{nj} + F^e_{im} S_{mn} {\cal T}_{jnkl} - F^e_{im} S_{mn} F^{e^T}_{nj} ({\cal T}_{pqkl} F^{e^{-1}}_{qp} )\right] . ~~~~~
\end{equation*}

\end{enumerate}

%%%%% Section 4 %%%%%

\section{Uniaxial compression of copper: model results}
\label{sec:4}

\subsection{Parameter selection and evaluation}
\label{sec:4_1}

The uniaxial compression experimental results used for parameter evaluation in this work are those from \citet{Chen_2019} and \citet{Follansbee_1988} where they present the stress-strain curves of OFHC copper at multiple conditions of strain rate and initial temperature.

The parameters used in our computation are listed in Table \ref{tab:parameters}. Many of the material parameters for fcc copper, such as those that determine the elastic moduli and the heat capacity, are well known. In particular, we assume that the mass density $\bar{\rho}_M$ and specific heat capacity $c_p$ to be constants for the loading conditions described in this manuscript, and stipulate that the crystallographic elastic moduli vary linearly with temperature:
\begin{equation}
 C_{ij} (T) = C_{ij}^0 + m_{ij} T ,
\end{equation}
with $\{ij\} = \{11\}$, $\{12\}$, or $\{44\}$, which give the temperature-dependent single-crystal shear modulus according to the relationship
\begin{equation}\label{eq:mu}
 \mu_{\text{sc}} (T) = \sqrt{ C_{44} (T) \left( \dfrac{C_{11} (T) - C_{12} (T)}{2} \right) } .
\end{equation}
To account for the effect of grain boundaries on the shear stiffness, we empirically adjust the shear modulus of the polycrystal OFHC copper as follows:
\begin{equation}
 \mu (T) = \mu_{\text{sc}} (T) + \mu_1 ,
\end{equation}
where $\mu_1$ is a constant for our purposes. This is consistent with typical handbook and experimental data \citep[e.g.,][]{SimmonsBook_1971} that indicates that the shear modulus of the polycrystal aggregate is uniformly higher than $\mu_{\text{sc}} (T)$ computed in accordance with Eq.~\eqref{eq:mu}, with some uncertainty.

\begin{table}
\scriptsize
\begin{center}
\caption{\label{tab:parameters}List of parameters and initial conditions}
\begin{tabular}{lll}
\hline
Parameter & Definition or meaning & Value \\
\hline\hline
$\bar{\rho}_M$ & Mass density & 8960 kg m$^{-3}$ \\
$c_p$ & Specific heat capacity & 380 J kg$^{-1}$ K$^{-1}$ \\
$C_{11}^0$ & Elastic moduli parameter & 179.5 GPa \\
$m_{11}$ & Elastic moduli parameter & -36.3 MPa K$^{-1}$ \\
$C_{12}^0$ & Elastic moduli parameter & 126.4 GPa \\
$m_{12}$ & Elastic moduli parameter & -16.4 MPa K$^{-1}$ \\
$C_{44}^0$ & Elastic moduli parameter & 82.5 GPa \\
$m_{44}$ & Elastic moduli parameter & 25.7 MPa K$^{-1}$ \\
$\mu_1$ & Shear modulus adjustment & 6 GPa \\
$\alpha_T$ & Stress scale parameter & 2 \\
$b$ & Burgers vector & 0.257 nm \\
$a^{\text{self}}$ & Dislocation interaction coefficient & 0.122 \\
$a^{\text{copl}}$ & Dislocation interaction coefficient & 0.122 \\
$a^{\text{hirth}}$ & Dislocation interaction coefficient & 0.070 \\
$a^{\text{colli}}$ & Dislocation interaction coefficient & 0.625 \\
$a^{\text{gliss}}$ & Dislocation interaction coefficient & 0.137 \\
$a^{\text{lomer}}$ & Dislocation interaction coefficient & 0.122 \\
$k_c$ & Mean free path parameter & 12.0 \\
$k_{nc}$ & Mean free path parameter & 180.0 \\
$a$ & Atomic length scale & 5.14 nm \\
$(a/b) t_0$ & Atomic time scale & 1 ps \\
$T_P$ & Depinning barrier & $40 800$ K \\
$\chi_{\text{ss}}$ & Steady-state effective temperature (in units of $e_D$) & 0.25 \\
$\kappa_1^{\alpha}$ & Hardening parameter & 100.0 \\
$\kappa_{\chi}$ & Effective temperature increase rate & 6.0 \\
\hline
\end{tabular}
\end{center}
\end{table}

In this work we suppose that the dislocation interaction coefficients $a^{\alpha \beta}$ and $d^{\alpha \beta}$, which determine the Taylor stress $s_T^{\alpha}$ and the dislocation mean free path $l^{\alpha}$ respectively, are the same as in conventional literature \citep[e.g.,][]{kubin_2008,dequiedt_2015,bronkhorst_2019}. Specifically, in fcc copper, for which there are 12 slip systems and four slip planes, the set of quantities $a^{\alpha \beta}$ contains six independent coefficients as follows:
\begin{enumerate}
 \item $a^{\alpha \alpha} = a^{\text{self}}$ (self interaction) ;
 \item $a^{\alpha \beta} = a^{\text{copl}}$ if $\mathbf{m}^{\alpha} = \mathbf{m}^{\beta}$ and $\mathbf{s}^{\alpha} \neq \mathbf{s}^{\beta}$ (dipolar interaction) ;
 \item $a^{\alpha \beta} = a^{\text{hirth}}$ if $\mathbf{m}^{\alpha} \neq \mathbf{m}^{\beta}$ and $\mathbf{s}^{\alpha} \perp \mathbf{s}^{\beta}$ (Hirth lock) ;
 \item $a^{\alpha \beta} = a^{\text{colli}}$ if $\mathbf{m}^{\alpha} \neq \mathbf{m}^{\beta}$ and $\mathbf{s}^{\alpha} = \mathbf{s}^{\beta}$ (collinear interaction) ;
 \item $a^{\alpha \beta} = a^{\text{gliss}}$ if $\mathbf{m}^{\alpha} \neq \mathbf{m}^{\beta}$, $\mathbf{s}^{\alpha} \neq \mathbf{s}^{\beta}$ with one of the two slip directions in both slip planes (glissile junction) ;
 \item $a^{\alpha \beta} = a^{\text{lomer}}$ otherwise (Lomer lock)  .
\end{enumerate}
The mean free path coefficients are given by
\begin{equation}\label{eq:dab1}
 d^{\alpha \beta} = \dfrac{a^{\alpha \beta}}{k_c^2}
\end{equation}
for intersecting slip systems, and
\begin{equation}\label{eq:dab2}
 d^{\alpha \beta} = \dfrac{a^{\alpha \beta}}{k_{nc}^2}
\end{equation}
for self interaction and coplanar slip systems. There is uncertainty in the values of $k_c$ and $k_{nc}$ but it appears from the present study to have a mostly cosmetic effect on the stress-strain curves.

Other parameters need to be determined separately. While \citet{langer_2010} and \citet{langer_2015}, in their work on the isotropic version of the thermodynamic dislocation theory for which there is only one single dislocation density variable, provided the values of many of these parameters, there is no a priori reason why values of parameters that pertain to dislocation generation and motion should remain unaltered in going from one dislocation density to 12 dislocation densities on as many distinct slip systems, which give rise to complex interactions between dislocations. Scalar thermodynamic parameters, however, should not depend on whether we adopt an isotropic or crystal-plasticity description. Thus, the dislocation depinning barrier $T_P$, which was unambiguously determined in \citet{langer_2010} to be around $40 800$ K from the steady-state stress of the experimental data available to them, should remain the same in our present calculations, even though we do not have the luxury of clean steady-state stresses at high temperatures at our disposal. In the same spirit, the steady-state effective temperature $\chi_{\text{ss}}$, in units of the dislocation formation energy $e_D$, is taken to be 0.25, as in the literature on the scalar, isotropic version of the TDT theory, since the effective temperature describes the state of inherent structural disorder. We then adjust $\kappa_1^{\alpha}$ to fit the initial hardening rate; the ratio $a/b$ and the rate parameter $\kappa_{\chi}$, the ratio of length scales $a/b$, and the initial effective temperature are adjusted to fit the data at large strains. Importantly, there is no a priori reason why the distinct samples tested at different thermal temperatures $T$ must share the same initial dislocation densities and effective temperature. Importantly, the stress-strain behavior depends only on the product $(a/b) \alpha_T$ if the initial dislocation densities $\rho^{\alpha}$ adjusted accordingly, inversely proportional to the ratio $a/b$, with all other parameters and initial conditions held fixed. This freedom of parameter selection is addressed by choosing $a/b$, where $a$ is the minimum separation between dislocation lines on a given slip system, to be of ${\cal O} (10)$, while using reasonable values for the initial dislocation densities.

\begin{figure}
\begin{center}
\begin{subfigure}[b]{.45\textwidth}
\includegraphics[width=\textwidth]{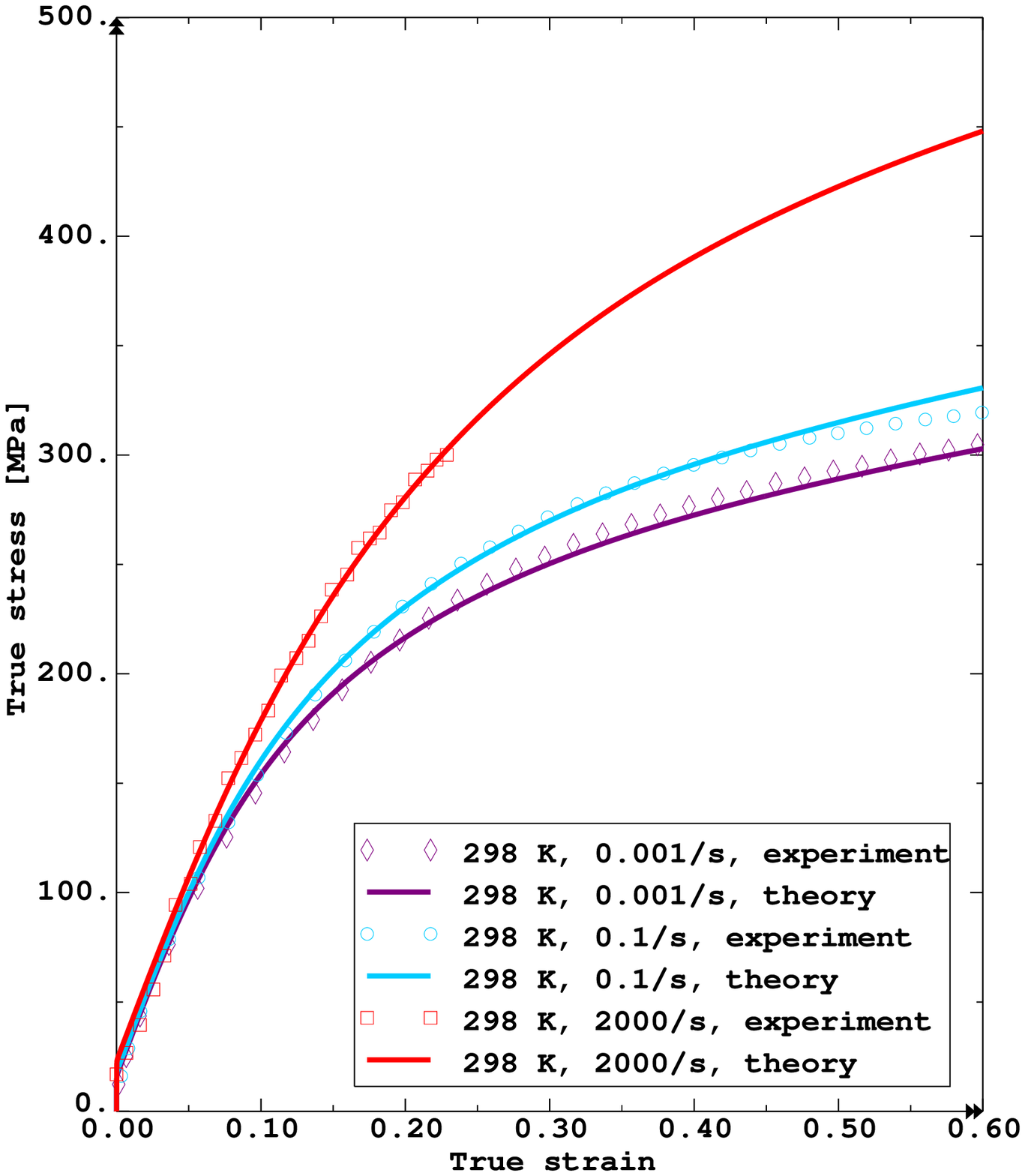}\caption{$T = 298$ K, various loading rates}
\end{subfigure}
\begin{subfigure}[b]{.45\textwidth}
\includegraphics[width=\textwidth]{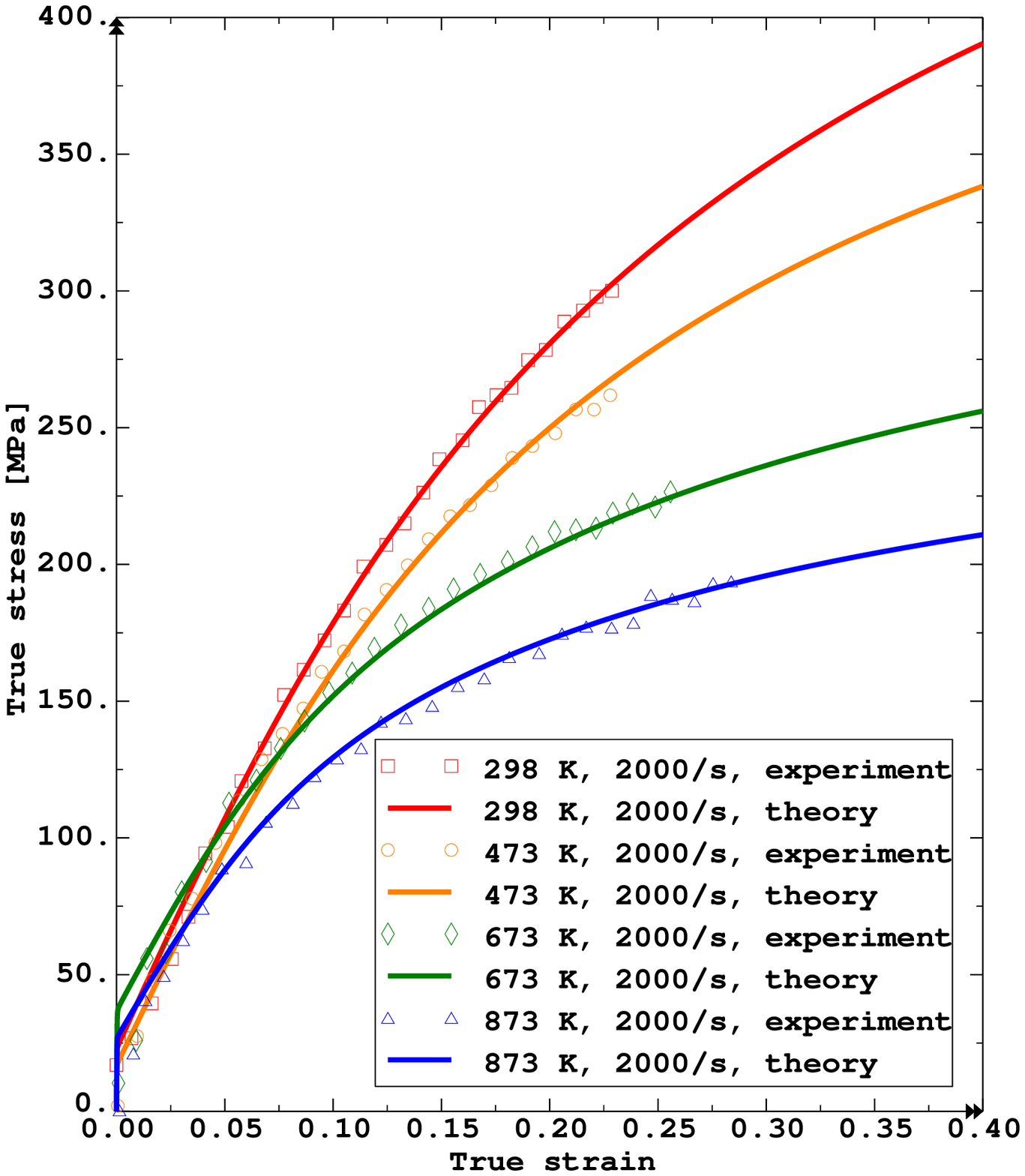}\caption{Various temperatures, 2000 s$^{-1}$}
\end{subfigure}
\caption{\label{fig:coppercube_ss}Stress-strain behavior of the millimeter-cube fcc copper polycrystal under uniaxial compression.}
\end{center}
\end{figure}

\subsection{Results}
\label{sec:4_2}

Simulations on the millimeter-cube fcc copper under uniaxial compression were performed, at a (true) strain rate of $10^{-3}$ s$^{-1}$ and $10^{-1}$ s$^{-1}$ for an initial temperature of $T (t = 0) = 298$ K, and at a strain rates of $2 \times 10^3$ s$^{-1}$ for temperatures $T (t = 0) = 298$, 473, 673, and 873 K. We assume isothermal deformation; the adiabatic assumption at the highest strain rate does not result in appreciable difference in the stress-strain response, in the strain regime for which experimental data is available to us. In other words, there is insufficient information for us to determine whether deformation is isothermal or adiabatic, or the value of the Taylor-Quinney coefficient. The initial conditions are listed in Table \ref{tab:initial}; grain orientations are determined from EBSD data. The results are shown in Fig.~\ref{fig:coppercube_ss}, demonstrating good agreement between theory and experiment.

\begin{table}
\scriptsize
\begin{center}
\caption{\label{tab:initial}List of initial dislocation densities $\rho^{\alpha}(t = 0)$ and effective temperature $\chi(t = 0)$ for uniaxial compression}
\begin{tabular}{llll}
\hline
Temperature (K) & Strain rate (s$^{-1}$) & $\rho^{\alpha}(t = 0)$ (mm$^{-2}$) & $\chi(t = 0)$ \\
\hline\hline
298 & 0.001 & $2 \times 10^5$ & 0.185 \\
298 & 0.1 & $2 \times 10^5$ & 0.185 \\
298 & 2000 & $2 \times 10^5$ & 0.195 \\
473 & 2000 & $2 \times 10^6$ & 0.21 \\
673 & 2000 & $2 \times 10^6$ & 0.21 \\
873 & 2000 & $2 \times 10^6$ & 0.225 \\
\hline
\end{tabular}
\end{center}
\end{table}

%\begin{figure}
%\begin{center}
%\includegraphics[width=0.6\textwidth]{fig_k20q_stress2.eps}
%\caption{\label{fig:splotm}Load-displacement curves computed using the finite element method with $h = 20$, 40, and 90 $\mu$m, compared to experimental measurements.}
%\end{center}
%\end{figure}

%\begin{figure}
%\begin{center}
%\includegraphics[width=0.6\textwidth]{fig_k44n_stress_ndrx2.eps}
%\caption{\label{fig:splot}Load-displacement curve computed using the finite element method with $h = 40 \mu$m, compared to experimental measurements. Also shown is the load-displacement curve computed for ``pseudo-steel'' that does not undergo dynamic recrystallization, and with otherwise identical material parameters, to indicate the crucial role of DRX in material softening.}
%\end{center}
%\end{figure}

\section{Uniaxial tension and simple shear}
\label{sec:5}

\begin{figure}
\begin{center}
\includegraphics[width=0.6\textwidth]{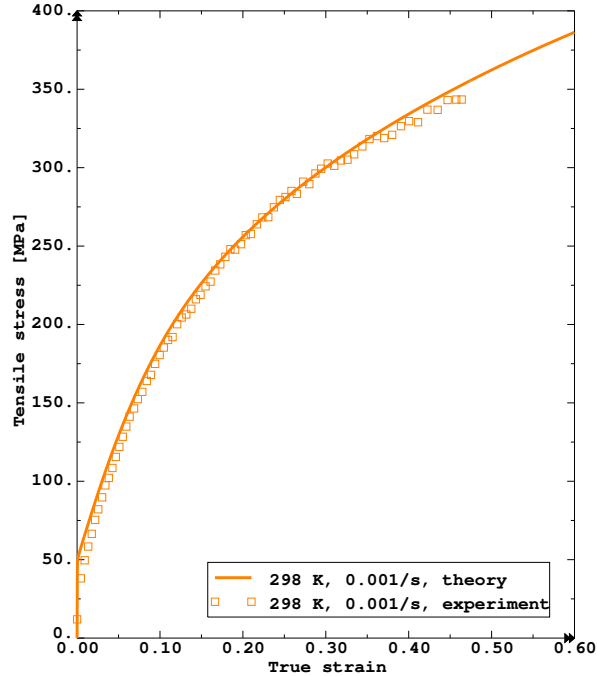}
\caption{\label{fig:coppercube_tension}Stress-strain curve for the copper polycrystal undergoing uniaxial tension, at a true strain rate of $10^{-3}$ s$^{-1}$ and temperature $T = 298$ K.}
\end{center}
\end{figure}

\begin{figure}
\begin{center}
\includegraphics[width=0.6\textwidth]{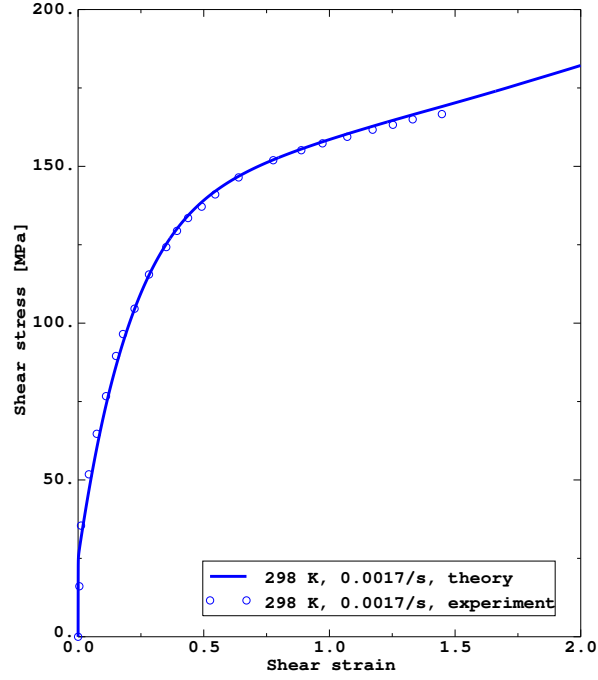}
\caption{\label{fig:coppercube_shear}Stress-strain curve for the copper polycrystal undergoing simple shear, at a strain rate of $1.7 \times 10^{-3}$ s$^{-1}$ and temperature $T = 298$ K.}
\end{center}
\end{figure}

To further validate the thermodynamic dislocation theory, we perform finite-element simulations, using the procedure described in Sec.~\ref{sec:3}, on the same polycrystal cube subjected to uniaxial tension and simple shear; we compare its stress-strain behavior with experiments described in \citet{bronkhorst_1992}. The results are shown in Figs.~\ref{fig:coppercube_tension} and \ref{fig:coppercube_shear} and demonstrate good agreement. There was no need for us to adjust any of the material parameters listed in Table \ref{tab:parameters}. However, we had to slightly adjust the initial dislocation densities and effective temperatures, as described in Table \ref{tab:initial_other}, in order to arrive at agreement at intermediate strains. This is not surprising. There is no reason a priori that material preparation for the \citet{bronkhorst_1992} experiments would result in the same dislocation densities, grain size, grain orientations, and state of configurational disorder, as in \citet{Follansbee_1988} and \citet{Chen_2019}; yet these initial conditions do influence the stress-strain response. That being said, both the initial hardening rate and the hardening rate at intermediate strains depend only on the material parameters in Table \ref{tab:parameters} and not on material preparation; thus the present agreement between theory and experiments provides a strong validation of the thermodynamic dislocation theory.

\begin{table}
\scriptsize
\begin{center}
\caption{\label{tab:initial_other}List of initial dislocation densities $\rho^{\alpha}(t = 0)$ and effective temperature $\chi(t = 0)$ for uniaxial tension and simple shear}
\begin{tabular}{lllll}
\hline
Experiment & Temperature (K) & Strain rate (s$^{-1}$) & $\rho^{\alpha}(t = 0)$ (mm$^{-2}$) & $\chi(t = 0)$ \\
\hline\hline
Uniaxial tension & 298 & 0.001 & $2 \times 10^6$ & 0.19 \\
Simple shear & 298 & 0.0017 & $2 \times 10^6$ & 0.18 \\
\hline
\end{tabular}
\end{center}
\end{table}

\section{Discussions}
\label{sec:6}

The thermodynamic dislocation theory described here is based on the premise that dislocations in a deforming crystal constitute material configurations that can, in principle, be enumerated. Thus, a system of dislocations correspond to some configurational energy and entropy and, therefore, a well-defined configurational or effective temperature. This effective temperature is ordinarily substantially higher than the thermal temperature, because thermal fluctuations alone can affect dislocation mobility but the energy represented is not great enough to create dislocations during deformation. This is then the foundation of thermally activated dislocation motion where motion is still dominated by plastic work put into the material. As a result, the configurational subsystem of the deforming crystal is only weakly coupled to the thermal subsystem, and is amenable to a thermodynamic description as described in this paper. 

\begin{figure}
\begin{center}
\includegraphics[width=0.6\textwidth]{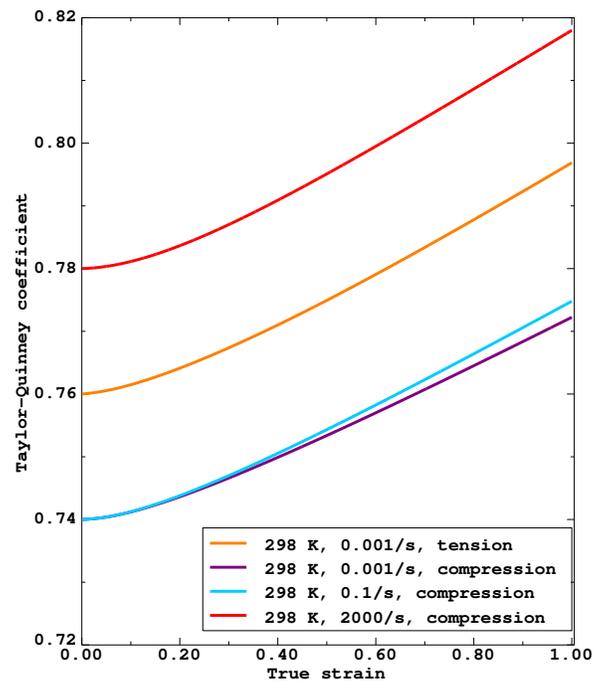}
\caption{\label{fig:tq_plot}Computed value of the Taylor-Quinney coefficient as a function of strain for tension and compression modes of loading at room temperature and varying rates of deformation.}
\end{center}
\end{figure}

Our thermodynamic description contains two crucial ingredients. The first of these is the direct connection between the dislocation densities $\rho^{\alpha}$ in the non-equilibrium steady state and the effective temperature $\chi$, as in Eq.~\eqref{eq:rho_ss0}; this is a consequence of the second law of thermodynamics, as is discussed earlier in the manuscript. The second crucial ingredient is the proportionality between the rate at which the dislocation densities evolve and the plastic work rate, as described by Eq.~\eqref{eq:rhodot}. This arises from the storage of input work in dislocation structures, and is an ingredient absent from most conventional continuum dislocation theories which are based largely on kinematics of dislocation motion only. These two physical ingredients constitute the most important elements of the TDT and make the theory thermodynamically consistent and also allow for inclusion of definitions of the Taylor-Quinney coefficient (Eq.~\eqref{eq:Tdot}) which then become inherently related to the complete balance of energy within the system. The current model has been used to compute the evolution of the Taylor-Quinney coefficient for the different loading and state conditions covered in this work and is given in Fig. ~\ref{fig:tq_plot}. As demonstrated the current theory illustrates an evolving value and differing slightly with loading rate. Recent experimental work also suggests that the value of this plastic power conversion factor is significantly less than 1.0 \citep{rittel_2012,rittel_2017,Kingstedt_2019}. Much more experimental work and parallel theoretical comparison with the experimental datasets is required to elucidate the evolution of material temperature with deformation. This will greatly facilitate our development of theories accounting for the energetics of dislocation motion and enable a path to understanding material damage under adverse loading conditions.

In going from the isotropic version of TDT \citep{langer_2010,langer_2015} to the crystal-plasticity version, we encounter a number of important questions that pertain to the crystalline, non-isotropic nature of the solid. The first of these is that with as many dislocation density variables $\rho^{\alpha}$ as the number of slip systems, can we quantify the interaction between different slip systems? For fcc copper, we have been able to fit experimental data by assuming, as in conventional dislocation theories, prescribed slip resistances and mean free paths that couple the different slip systems (Eqs.~\eqref{eq:taylorstress}, \eqref{eq:dab1}, and \eqref{eq:dab2}), and set the interaction energy $e_N$ between different slip systems to be equal to zero in Eq.~\eqref{eq:rho_ss0}, which in effect rules out latent hardening for the current study. As discussed within the introduction, this remains a very important area of research, to understand and quantify the nature of dislocation interactions and their observed formation of intragranular dislocation substructures as the physical root of observed material hardening. This dislocation structure evolution has important implications for material damage as the heterogeneous stress field will evolve with deformation. The theory proposed here as well as other recent work \citep{dequiedt_2015,Grilli_2018,Arora_2019} offer continuum representation of discrete dislocation processes which are however able to access comparisons with large experimental databases for many different material types. Advances to these continuum theories have already benefited from discrete dislocation dynamics simulations (e.g. \citep{Madec_2002,Madec_2003,Devincre_2006,Devincre_2008}) to quantify the differing interactions within the fcc crystal structure. This relationship between available experimental information, physics simulations of dislocation mechanics, and continuum theories is integral to making tangible progress. Although discrete dislocation dynamics generally lacks accounting for thermodynamic quantities, it is continuing to be used effectively for the study of dislocation mechanics \citep{Rao_2019,Cho_2018,Santos-Guemes_2018}. New high-resolution continuum and phase field techniques hold promise for computing dislocation interaction kinetics with some able to do so in a thermodynamically consistent way \citep{Acharya_2004,Acharya_2006,Acharya_2010,Peng_2020,Eghtesad_2018,Arora_2019,Zeng_2016}. The modeling community is most advanced for fcc systems with significant amounts of work remaining to quantify slip for other common bcc and hcp crystal systems given that these systems are generally more complex.

\section{Concluding remarks}
\label{sec:7}

In this paper, we formulated a single crystal, finite-deformation version of the thermodynamic dislocation theory, first pioneered by Langer, Bouchbinder, and Lookman~\citep{langer_2010,langer_2015} in the language of isotropic plasticity and infinitesimal deformation. With only a small handful of adjustable parameters, finite-element simulations demonstrate favorable agreement between theory and a series of experiments on polycrystal fcc copper strained under a range of temperatures, loading rates, and loading configurations including uniaxial compression, tension, and simple shear. Our work constitutes an essential first step of this line of thought towards a physics-based, thermodynamically-consistent description of dislocation plasticity in arbitrary geometries and loading configurations. The topic of hardening and complex dislocation interactions represented via a continuum theory is introduced into this thermodynamic framework. The introduction of plastic work as an energy driver for the production of additional defects introduces the concept of the energy of interactions and the potential of interacting stress fields around the cores of dislocations as a factor in material hardening which will be explored further.

\section*{Acknowledgements}

CL would like to thank J. S. Langer, K. C. Le, and R. A. Lebensohn for insightful discussions. CL was partially supported by the Center for Nonlinear Studies at the Los Alamos National Laboratory over the duration of this work. CB acknowledges support from the University of Wisconsin Alumni Research Foundation. All authors were partially supported by the DOE/DOD Joint Munitions Program and and LANL LDRD Program Project 20170033DR. The authors declare no conflicts of interest. 

%% The Appendices part is started with the command \appendix;
%% appendix sections are then done as normal sections
%% \appendix

%% \section{}
%% \label{}

\section*{References}

%% If you have bibdatabase file and want bibtex to generate the
%% bibitems, please use
%%
\bibliographystyle{elsarticle-harv} 
\bibliography{jmps_copper_05_CAB}

%% else use the following coding to input the bibitems directly in the
%% TeX file.

%\begin{thebibliography}{00}

%% \bibitem{label}
%% Text of bibliographic item

%\bibitem{}

%\end{thebibliography}
\end{document}